# NORMAL MODE TUNES FOR LINEAR COUPLED MOTION IN SIX DIMENSIONAL PHASE SPACE*

G. Parzen, Brookhaven National Laboratory, Upton, NY 11973 USA

## Abstract

The motion of a particle in 6-dimensional phase space in the presence of linear coupling can be written as the sum of 3 normal modes. A cubic equation is found for the tune of the normal modes, which allows the normal mode tunes to be computed from the 6 × 6 one turn transfer matrix. This result is similar to the quadratic equation found for the normal mode tunes for the motion of a particle in 4-dimensional phase space. These results are useful in tracking programs where the one turn transfer matrix can be computed by multiplying the transfer matrices of each element of the lattice. The tunes of the 3 normal modes for motion in 6-dimensional phase space can then be found by solving the cubic equation. Explicit solutions of the cubic equation for the tune are given in terms of the elements of the 6 × 6 one turn transfer matrix.

## I. INTRODUCTION

The motion of a particle in 6-dimensional phase space in the presence of linear coupling can be written as the sum of 3 normal mode. A cubic equation is found for the tune of the normal modes, which allows the normal mode tune to be computed from the 6 × 6 one turn transfer matrix.

This result is similar to the quadratic equation [1] found for the normal mode tune for the motion of a particle in 4-dimensional phase space. These results are useful in tracking programs where the one turn transfer matrix can be computed by multiplying the transfer matrices of each element of the lattice. The tune of the 3 normal modes for motion in 6-dimensional phase space can then be found by solving the cubic equation. Explicit solutions of the cubic equation for the tune are given in terms of the elements of the 6 × 6 one turn transfer matrix.

## II. A CUBIC EQUATION FOR NORMAL MODE TUNES

The particle coordinates are assumed to be $x$, $p_x$, $y$, $p_y$, $z$, $p_z$. Then the linear motion of the particle about some central orbit can be described by a 6 × 6 transfer matrix $T(s, s_0)$.

$$x(s) = T(s, s_0) x(s_0)$$

$$x = \begin{bmatrix} x \\ p_x \\ y \\ p_y \\ z \\ p_z \end{bmatrix} \quad (1)$$

Note that the symbol $x$ is used to denote the column vector $x$ and also the first element of this column vector. The meaning of $x$ should be clear from the context. This notation is used in several places in this paper.

It is assumed that the equations of motions can be derived from a hamiltonian, and the matrix T is symplectic:

$$\begin{aligned} T\overline{T} &= I \\ \overline{T} &= \tilde{S}\,\tilde{T}\,S \end{aligned} \quad (2)$$

$\tilde{T}$ is the transpose of T.

$I$ is the 6 × 6 identity matrix, and $S$ can be written in terms of 2 × 2 matrices as

$$S = \begin{bmatrix} S & 0 & 0 \\ 0 & S & 0 \\ 0 & 0 & S \end{bmatrix} \quad (3)$$

The 2 × 2 matrix, also called $S$, is given by

$$S = \begin{pmatrix} 0 & 1 \\ -1 & 0 \end{pmatrix} \quad (4)$$

It is convenient to also write T in terms of 2 × 2 matrices

$$T = \begin{bmatrix} T_{11} & T_{12} & T_{13} \\ T_{21} & T_{22} & T_{23} \\ T_{31} & T_{32} & T_{33} \end{bmatrix} \quad (5)$$

One may note that the 2 × 2 matrices of $\overline{T}$ are $(\overline{T})_{ij} = \overline{T}_{ji}$. The eigenfunctions of the one turn transfer matrix $T(s + L, s)$, where $L$ is the period of the forces acting on the particle, obey

$$T(s + L, s)x = \lambda x$$

Because T is symplectic then if $\lambda$ is an eigenvalue, $1/\lambda$ is also an eigenvalue [1,2] and the 6 eigenvalue can be arranged in 3 pairs of $\lambda$ and $1/\lambda$. For stable motion, $|\lambda| = 1$ and $\lambda$ can be written as

$$\begin{aligned} \lambda &= \exp(i\mu) \\ \mu &= 2\pi\nu, \end{aligned} \quad (6)$$

where it is assumed that the period is one turn.

To find the tunes of the normal modes, one has to find the eigenvalues of T, $\lambda$, which are given by $|T - \lambda I| = 0$, where $|T|$ indicates the determinant of T. It is more convenient to find the eigenvalues of the matrix C defined by

$$C = \frac{1}{2}(T + \overline{T}) \quad (7)$$

The eigenvalues of C, Λ, are

$$\Lambda = \frac{1}{2}(\lambda + 1/\lambda),$$
$$\Lambda = \cos\mu \quad (8)$$

and the Λ are determined by

$$|C - \Lambda I| = 0 \quad (9)$$

It is convenient to write C in terms of $2 \times 2$ matrices. Note that the $2 \times 2$ elements of C are

$$C_{ij} = (T_{ij} + \overline{T}_{ji})/2 \quad (10)$$

and thus

$$C_{ii} = \frac{1}{2}Tr(T_{ii})I$$
$$C_{ij} = \overline{C}_{ji} \quad (11)$$

We can then write C as

$$C = \begin{bmatrix} t_1 I & C_{12} & C_{13} \\ C_{21} & t_2 I & C_{23} \\ C_{31} & C_{32} & t_3 I \end{bmatrix} \quad (12)$$

$$t_i = \frac{1}{2}T_r(T_{ii})$$

To evaluate $|C - \Lambda I|$, $C - \Lambda I$ will be multiplied from the left by a sequence of matrices, each with determinant 1, to produce a matrix which is upper triangular; that is, the elements of this matrix below the diagonal are all zero. This triangular matrix has the same determinant as $C - \Lambda I$, and its determinant can be found by multiplying all the diagonal elements of this matrix.

First, multiply $C - \Lambda I$ from the left by the matrix

$$\begin{bmatrix} 1 & 0 & 0 \\ -C_{21}/\overline{t}_1 & 1 & 0 \\ 0 & 0 & 1 \end{bmatrix} \quad (13)$$

$$\overline{t}_1 = t_1 - \Lambda$$

This matrix is designed to eliminate $C_{21}$ and the multiplication gives the result

$$\begin{bmatrix} \overline{t}_1 I & C_{12} & C_{13} \\ 0 & (\overline{t}_2 - |C_{12}|/\overline{t}_1)I & C_{23} - C_{21}C_{13}/\overline{t}_1 \\ C_{31} & C_{32} & \overline{t}_3 \end{bmatrix} \quad (14)$$

The result $C_{12}\overline{C}_{12} = |C_{12}|$ has been used. Note that $C_{21} = \overline{C}_{12}$. Now multiply the matrix Eq. (14) by the matrix

$$\begin{bmatrix} 1 & 0 & 0 \\ 0 & 1 & 0 \\ -C_{31}/\overline{t}_1 & 0 & 1 \end{bmatrix} \quad (15)$$

which replaces the third row in Eq. (14) by the third row plus $-C_{31}/\overline{t}_1$ times the first row giving

$$\begin{bmatrix} t_1 I & C_{12} & C_{13} \\ 0 & (\overline{t}_2 - |C_{12}|/\overline{t}_1)I & C_{23} - C_{21}C_{13}/\overline{t}_1 \\ 0 & C_{32} - C_{31}C_{12}/\overline{t}_1 & (\overline{t}_3 - |C_{13}|/\overline{t}_1)I \end{bmatrix} \quad (16)$$

One more matrix multiplication is required to reduce $C - \Lambda I$ to an upper triangular matrix. Multiply matrix Eq. (16) by the matrix

$$\begin{bmatrix} 1 & 0 & 0 \\ 0 & 1 & 0 \\ 0 & -(C_{32} - C_{31}C_{12}/\overline{t}_1)/(\overline{t}_2 - |C_{12}|\overline{t}_1) & 0 \end{bmatrix} \quad (17)$$

This gives

$$\begin{bmatrix} \overline{t}_1 I & C_{12} & C_{13} \\ 0 & (\overline{t}_2 - |C_{12}|/\overline{t}_1)I & C_{23} - C_{21}C_{13}/\overline{t}_1 \\ 0 & 0 & \widetilde{t}_3 I \end{bmatrix} \quad (18)$$

$$\widetilde{t}_3 = [\overline{t}_3 - |C_{13}|/\overline{t}_1 - |C_{23} - C_{21}C_{13}/\overline{t}_1|/(\overline{t}_2 - |C_{12}|/\overline{t}_1)]$$

The matrix Eq. (18) is a triangular matrix with zero elements below the diagonal and its determinant, and thus $|C - \Lambda I|$, can be computed by multiplying the diagonal elements which gives

$$|C - \Lambda I|^{1/2} = \overline{t}_1(\overline{t}_2 - |C_{12}|/\overline{t}_1)[\overline{t}_3 - |C_{13}|/\overline{t}_1$$
$$- |C_{23} - C_{21}C_{13}/\overline{t}_1|/(\overline{t}_2 - |C_{12}|/\overline{t}_1)]$$
$$|C - \Lambda I|^{1/2} = \overline{t}_1[(\overline{t}_2 - |C_{12}|/\overline{t}_1)(\overline{t}_3 - |C_{13}|/\overline{t}_1)$$
$$- |C_{23} - C_{21}C_{13}/\overline{t}_1|] \quad (19)$$
$$|C - \Lambda I|^{1/2} = (1/\overline{t}_1)[(\overline{t}_1\overline{t}_2 - |C_{12}|)(\overline{t}_1\overline{t}_3 - |C_{13}|)$$
$$- |C_{23}\overline{t}_1 - C_{21}C_{13}|]$$

To simplify Eq. (19), note the following result

$$|C_{23}\overline{t}_1 - C_{21}C_{13}|I = (C_{23}\overline{t}_1 - C_{21}C_{13})(\overline{C}_{23}\overline{t}_1 - \overline{C}_{13}\overline{C}_{21})$$
$$= \left[|C_{23}|\overline{t}_1^2 - T_r(C_{23}\overline{C}_{13}\overline{C}_{21})\overline{t}_1 + |C_{21}||C_{13}|\right]I \quad (20)$$

which gives

$$|C - \Lambda I|^{1/2} = \overline{t}_1\overline{t}_2\overline{t}_3 - \overline{t}_2|C_{13}| - \overline{t}_3|C_{12}| - \overline{t}_1|C_{23}| + T_r(C_{23}C_{31}C_{12}) \quad (21)$$

Finally, the normal modes tunes, which are given by $|C - \Lambda I| = 0$ are determined by the cubic equation

$$(\Lambda - t_1)(\Lambda - t_2)(\Lambda - t_3) - (\Lambda - t_1)|C_{23}| - (\Lambda - t_2)|C_{31}| - (\Lambda - t_3)|C_{12}|$$
$$- T_r(C_{12}C_{23}C_{31}) = 0$$

$$\Lambda = \cos\mu, \quad \mu = 2\pi\nu$$

$$t_1 = \frac{1}{2}T_r(T_{11}), \quad t_2 = \frac{1}{2}T_r(t_{22}), \quad t_3 = \frac{1}{2}T_r(T_{33}) \quad (22)$$

$$C = \frac{1}{2}(T + \overline{T})$$

$$C_{ij} = \frac{1}{2}(T_{ij} + \overline{T}_{ji})$$

The $T_{ij}$ are the $2 \times 2$ elements of T and $C_{ij}$ are the $2 \times 2$ elements of C.

Equation (22) is a cubic equation, and the 3 roots of this equation $\Lambda_i = \cos\mu_i, i = 1,3$ will give the 3 tunes of the normal modes, $\nu_i = \mu_i/2\pi$. If $\mu_i$ is real the mode is stable; if $\mu$ has an imaginary part then the mode may be unstable.

To get a result for 4-dimensional coupled motion, we put $C_{23} = C_{31} = 0$ and Eq. (22) then gives

$$(\Lambda - t_1)(\Delta - t_2) - |C_{12}| = 0$$

$$C_{12} = \frac{1}{2}(T_{12} + \overline{T}_{21}), \ t_1 = \frac{1}{2}T_r(T_{11}), \ t_2 = \frac{1}{2}t_r(T_{22}) \quad (23)$$

Eq. (23) is the known result [1,3] for the tunes of the 2 normal modes for coupled motion in 4-dimensional phase space.

In the 4-dimensional case, Eq. (23) can be solved to find the two $\cos\mu_i$ of the normal modes. This gives the result for $\cos\mu$

$$\begin{aligned}
\cos\mu &= \frac{1}{2}(t_1 + t_2) \pm \frac{1}{2}\left[(t_1 - t_2)^2 + 4|C_{12}|\right]^{1/2} \\
t_1 &= \frac{1}{2}T_r(T_{11}), \ t_2 = \frac{1}{2}T_r(T_{22}) \quad (24) \\
|C_{12}| &= |T_{12} + \overline{T}_{21}|/4
\end{aligned}$$

where $T_{ij}$ are the 2 × 2 elements of the one turn transfer matrix. If one computes T by multiplying the transfer matrices of all the elements in the accelerator, then one can find $\cos\mu$, and the $\nu$ values, using Eq. (24).

Equation (24) is useful in tracking programs for finding the normal mode tunes, from the one turn transfer matrix, for coupled motion in 4-dimensional phase space.

A similar result can be found for coupled motion in 6-dimensional phase space by solving the cubic equation, Eq. (22).

Equation (22) can be written as

$$\begin{aligned}
\Lambda^3 &+ a_2\Lambda^2 + a_1\Lambda + a_0 = 0 \\
a_2 &= -(t_1 + t_2 + t_3) \\
a_1 &= t_1t_2 + t_2t_3 + t_3t_1 - |C_{12}| - |C_{23}| - |C_{31}| \\
a_0 &= -t_1t_2t_3 - T_r(C_{12}C_{23}C_{31}) + t_1|C_{23}| + t_2|C_{31}| + t_3|C_{12}| \\
t_1 &= \frac{1}{2}T_r(T_{11}), \ t_2 = \frac{1}{2}T_r(T_{22}), \ t_3 = \frac{1}{2}T_r(T_{33}) \\
C_{ij} &= \frac{1}{2}(T_{ij} + \overline{T}_{ji}) \quad (25)
\end{aligned}$$

The solutions of Eq. (25) when the 3 roots are all real, can be written as [4]

$$\begin{aligned}
\cos\mu_i &= (t_1 + t_2 + t_3)/3 + 2(r^2 + b^2)^{1/6}\cos(\alpha/3 + \delta_i) \\
\delta_i &= 0, 2\pi/3, -2\pi/3 \\
b &= |r^2 + q^3|^{1/2} \\
r &= \frac{1}{6}(a_1a_2 - 3a_0) - \frac{1}{27}a_2^3 \quad (26) \\
q &= \frac{1}{3}a_1 - \frac{1}{9}a_2^2 \\
\tan\alpha &= b/r
\end{aligned}$$

Note that the 3 values of $\delta_i, \delta_i = 0, 2\pi/3, = 2\pi/3$ will give the 3 roots from Eq. (26).

The condition for the 3 roots to be real is [4]

$$q^3 + r^2 \leq 0 \quad (27)$$

If $(q^3 + r^2) > 0$, then 2 roots are imaginary and the motion may be unstable. In order to get $q^3 + r^2 \leq 0$ and stable motion, one has to have $q < 0$.

Equation (26) can be used in a tracking program to find the 3 normal mode tunes from the one turn transfer matrix. The one turn transfer matrix can be computed by multiplying the transfer matrices for each element in the lattice.